\documentclass[pra,floatfix,preprint,nofootinbib,eqsecnum,12pt]{revtex4}
\usepackage{epsfig}
\usepackage{bm}
\usepackage{amsmath}
\usepackage{enumitem}
\usepackage{color}
\input epsf

\def\pfp{p^{(1p)}}
\def\be{\begin{equation}}
\def\ee{\end{equation}}
\def\bigstar{\mbox{\Large $*$}}

\def\HP{$(H,\Psi)$}
\def\HPs{$(H,\Psi) \ $}
\def\p0{\phi_0}

\def\pfp{{p^{(1p)}}}

\def\Dge{D^{\ge 1}}
\def\sk{\vskip .1in}

\def\be{\begin{equation}}
\def\ee{\end{equation}}

\begin{document}
\vspace{1cm}

\title{Quantum Multiverses\footnote{A pedagogical essay.}}
\author{James B.~Hartle}

\email{hartle@physics.ucsb.edu}

\affiliation{Santa Fe Institute, Santa Fe, NM 87501}
\affiliation{Department of Physics, University of California,Santa Barbara, CA 93106-9530}

\date{\today}

\begin{abstract}

A quantum theory of the universe consists of a theory of its quantum dynamics $(H)$ and a theory of its quantum state ($\Psi$). The theory $(H,\Psi)$ predicts quantum multiverses in the form of decoherent sets of alternative histories describing the evolution of the universe's spacetime geometry and matter  content.   A small part of one of these histories is observed by us. These consequences follow: (a) The universe generally exhibits different  quantum multiverses at different levels and kinds of coarse graining. (b) Quantum multiverses are not a choice or an assumption but are consequences of  \HP\ or not. (c) Quantum multiverses are generic for simple \HP. (d) Anthropic selection is automatic because observers are physical systems within the universe not somehow outside it. (e) Quantum multiverses can provide different mechanisms for  the variation constants in effective theories (like the cosmological constant) enabling anthropic selection. (f) Different levels of coarse grained multiverses provide different routes to calculation as a consequence of decoherence. We support these conclusions by analyzing the quantum multiverses of a variety of quantum cosmological models aimed at the prediction of observable properties of our universe. In particular we  show how the example of a multiverse  consisting of a vast classical  spacetime containing many  pocket universes having different values of the fundamental constants arises automatically as part of a quantum multiverse describing an eternally inflating false vacuum that decays by the quantum nucleation of true vacuum bubbles. In a FAQ we argue that the quantum multiverses of the universe are scientific, real, testable, falsifiable, and similar to those in other areas of science even if they are not directly observable on arbitrarily large scales. 
\end{abstract}


\maketitle

\tableofcontents

\bibliographystyle{unsrt}


\section{Introduction}
\label{intro}

The universe may present us with an ensemble of alternative possible situations only one of which is observed by us. In this paper we call such an ensemble a `multiverse'. 

A much discussed  classical example of a multiverse is a single, vast  cosmological spacetime containing  `pocket universes' at different locations. Physics inside different pockets is assumed to be governed by different low energy effective theories. For example, the effective theories could differ in the value of the cosmological constant $\Lambda$. This is a multiverse of pockets with different values of $\Lambda$,  only one which is observed by us --- the one we live in. For short,  we call this a {\it pocket multiverse}.  Only small values $\Lambda \lesssim 10^{-120}$ are consistent with the rest of our cosmological data including a description of us as physical systems within the universe and the formation of galaxies by the present age  (e.g.\cite{BT86,Wei89,HH13}). Had we not yet measured $\Lambda$ we would predict that our pocket has a value in this small range\footnote{Some might restrict the term `multiverse' to just this example, but in our opinion there is clarity and simplicity in our more general definition.}. This is a simple example of {\it anthropic selection}. We won't observe properties of a universe where we cannot exist.  We will discuss this essentially classical  example in a quantum mechanical context in in Section \ref{bubbles}.

Quantum theories of a closed system like the universe provide  multiverses in the form of decoherent sets of alternative coarse-grained histories of the universe\footnote{In much other work we have called decoherent sets of alternative coarse grained histories `realms'. In this paper we use `multiverse'. }. Such sets are are quantum multiverses in the sense of the first sentence in this paper --- an ensemble of possible histories only one of which is observed by us. Quantum mechanics predicts probabilities for which of  the individual members of the ensemble of histories happens starting from from theories of the universe's dynamics ($H$) and quantum state ($\Psi$). When these probabilities are conditioned on a description of our observational situation (including us) we get probabilities for what we observe of the universe. 
 
 Quantum multiverses have been discussed extensively in connection with the fundamentals of quantum mechanics\footnote{See for example the discussions  in \cite{mw10,Deu11,Wal12}. }. 
This essay is devoted to discussing  quantum multiverses in the context of quantum cosmology. We will illustrate the idea in  simple concrete, calculable models  based on the author's joint work with Stephen Hawking and Thomas Hertog calculating the predictions for observations of the no-boundary wave function proposal \cite{HH83} for $\Psi$ (mainly \cite{HH15b,HH13,HHH10b,Her14}). We shall show how  predicting the results of our observations is enabled by quantum mechanics.  In particular, we shall show how a pocket multiverse emerges from an appropriate \HP. 

These models support the following conclusions about quantum multiverses in general:
\sk

{\it a. Many Quantum Multiverses.} The theory \HP\  does not  predict just one multiverse of alternative histories. It predicts many different multiverses at different levels and kinds of coarse graining. All are available for prediction. 
\sk

{\it b. Quantum multiverses are Not a Choice.} Quantum multiverses are not a choice or an assumption separate from the theory \HP. Rather they follow or do not follow from \HP. 
\sk

{\it c. Quantum Multiverses are Generic.} Simple, manageable, discoverable  theories  \HP\ generically predict quantum multiverses with many histories.  To predict just one history with  certainty  all of present complexity would have to be encoded \HP. Rather we expect that the  that present complexity arose  not just from  \HP, but  through a multiverse of histories that describe frozen accidents that occurred over the course of  the universe's history --- chance events  that could have gone one way or the other for which the consequences of the way it did go proliferated. The accidents of biological evolution are an example.
\sk

{\it d. Anthropic Selection is Automatic.} Anthropic selection is an automatic consequence of quantum mechanical probabilities for observations. This because observers are physical systems within the universe not somehow outside it. Probabilities for our observations are conditioned on a description of our observational situation and we won't observe what is where we cannot exist.  Anthropic reasoning does not rely on some anthropic principle and is not a choice to be be made or not made. It is an automatic consequence of calculating probabilities for observations. 
\sk

{\it e. Quantum multiverses provide several mechanisms for the constants of effective theories to vary.}  The pocket universe considered above concerns one history with $\Lambda$  varying from place to place.  But there are \HPs leading to multiverses of spatially homogeneous histories with $\Lambda$ the same at all places in each history but differing from history to history also enabling anthropic selection. We will provide examples of both mechanisms in Sections \ref{bubbles} and \ref{lambda}.
\sk

{\it f. Two Routes to Coarse-Graining.}  Quantum multiverses are restricted to sets of alternative coarse grained histories that are decoherent --- that is, restricted to  sets that have negligible quantum interference between the individual histories in the set  as a consequence of \HP. This ensures that the probabilities are consistent with the rules of probability theory \cite{classicDH}. Further coarse graining can be carried out either by summing probabilities or by summing quantum amplitudes. That can be a considerable computational advantage as we illustrate by example in Section \ref{bubbles}. 
\sk

 Understanding multiverses as decoherent sets of alternative coarse-grained histories of a quantum universe can help address some of the concerns, objections, that have been raised about multiverses. Multiverses arise naturally and inevitably in quantum theory as well as many other areas of science.  The author's views on some of the concerns are given in the FAQ in Appendix \ref{discussion}. 

The paper is structured as follows:   Section \ref{dh} uses two-slit model closed systems to describe the decoherent (or consistent) histories formulation of quantum mechanics (DH) which we use throughout. Sections III -VIII present the five model quantum cosmologies  and their quantum multiverses on which we base our conclusions above. Conclusions are in Section \ref{concl}.  The FAQ in Appendix \ref{discussion}  briefly addresses some of the concerns and objections that have been raised concerning multiverses.  A further appendix \ref{lm} gives more details about DH beyond those in Section \ref{dh}. 

\section{Quantum Mechanics for the Universe Illustrated in Two-Slit Models}
\label{dh}
\label{2slit-mods}

This section uses three  models based on the two-slit experiment to illustrate how different quantum multiverses  can arise in the same physical situation illustrating points (a)-(d) of the Introduction. 
We assume the decoherent or consistent histories formulation of the quantum mechanics of a closed system  such as the universe\footnote{For classic references see \cite{classicDH}. For a tutorial see, e.g. \cite{Har93a}.}.
Decoherent histories quantum theory (DH)  is logically consistent, consistent with experiment as far as is known, consistent with the rest of modern physics such as special relativity and quantum field theory, general enough for histories, general enough for cosmology, and generalizable to include semiclassical quantum gravity. Copenhagen quantum mechanics is contained within DH as an approximation appropriate for measurement situations. DH can be thought of as an extension, clarification, and, to some extent,  a completion of the program started by Everett \cite{Eve57}. DH may not be the only formulation of quantum mechanics with these properties but it is the only one we have at present.

\begin{figure}[t]
\includegraphics[height=2.2in]{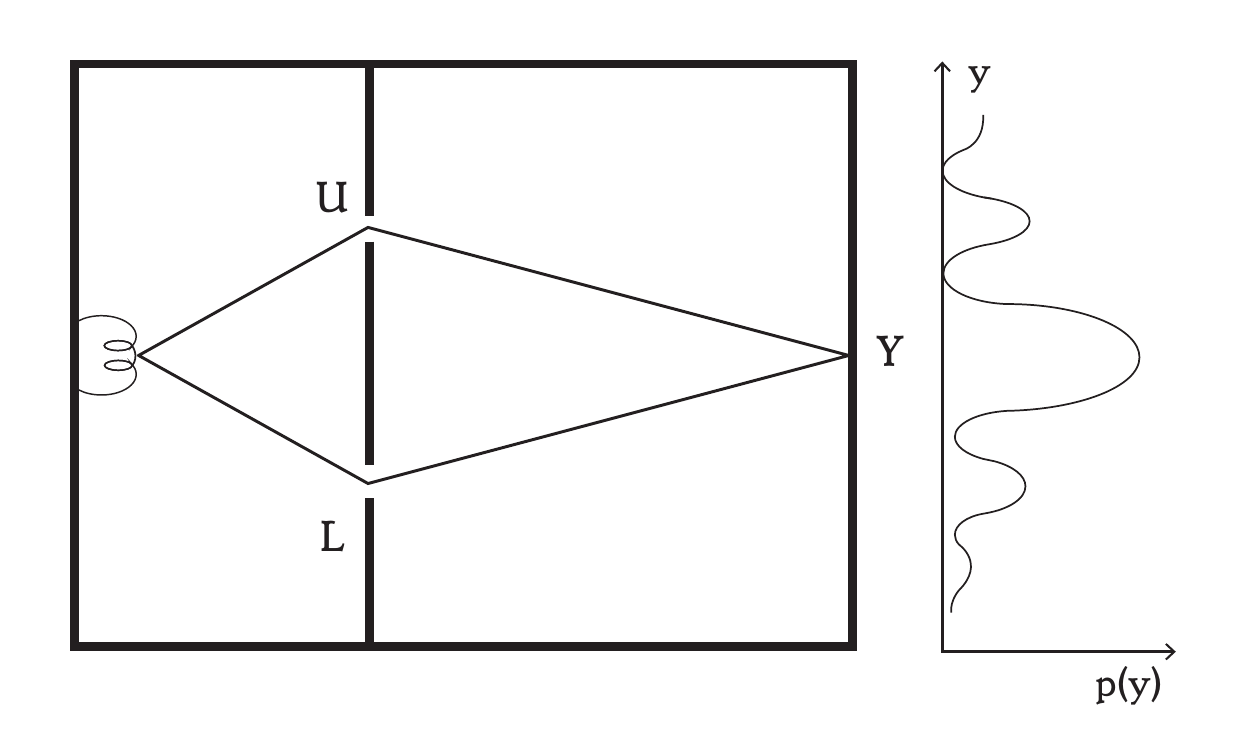}  \\ \hfill  \\
\includegraphics[height=2.2in]{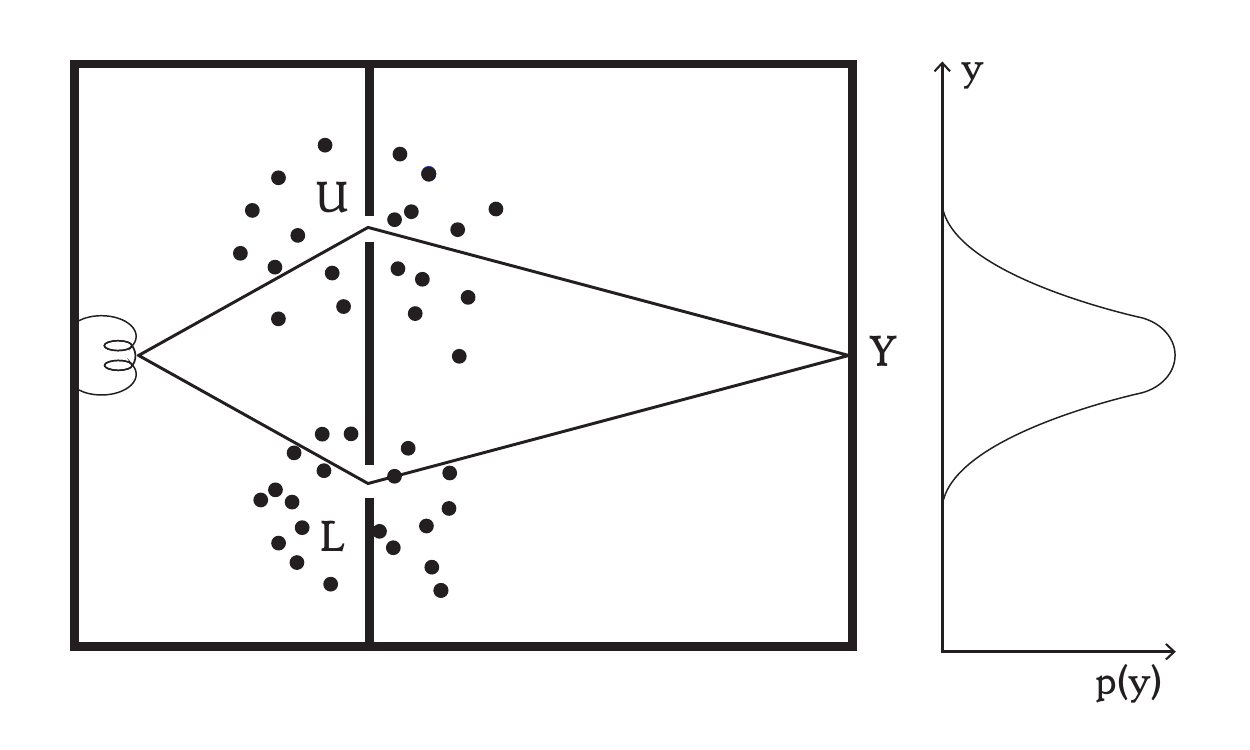} \hfill \\
\includegraphics[height=2.2in]{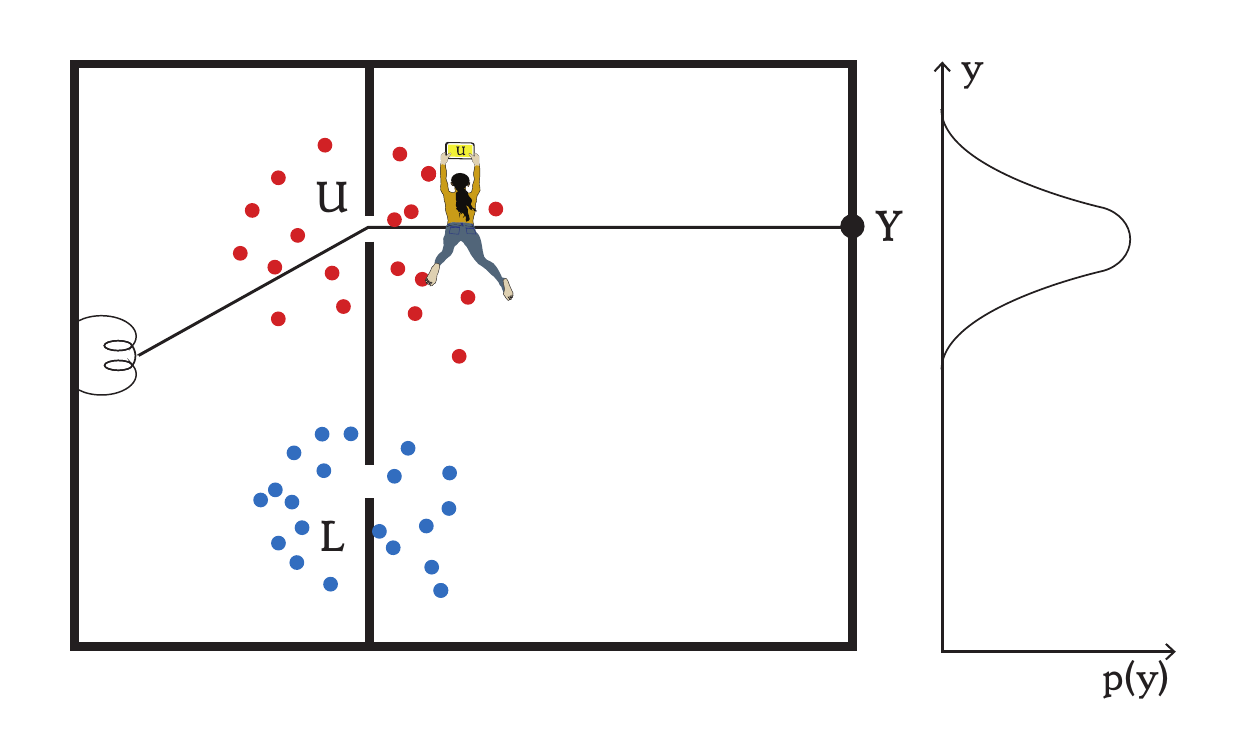}
\caption{Three idealized model closed systems based on  the two-slit setup. The top model (TSS)  is just the familiar two-slit example. The second model  (TSG) includes a gas of particles in the vicinity of the slits that scatter weakly off the electron as it passes through. The third model (TSGO) considers a very large box of this kind with an observer who can pass through the slits (Figure \ref{fig2}). The gas  particles have a distinguishable color detectable by the observer --- red near the upper slit and blue near the lower one. 
The probabilities for detection in an interval $Y$ on the far screen is indicated schematically in the graphs on the right. }
\label{2slit}
\end{figure}

The basic ideas of DH that will be needed in this paper can be introduced with three model two-slit situations each in a closed box --- three very simple model universes. We will illustrate a variety of quantum multiverses with these simple situations. 
The  three models are illustrated in Figure \ref{2slit}. We discuss these informally with a minimum of equations in this section. The same discussion with more equations can be found in Appendix \ref{lm}.

{\it TSS:} In the first model  the box contains  an electron gun at left that emits an electron which moves through a screen with two slits $S=(U,L)$ to arrive at a further screen at right in one of a set of position intervals $Y=(1,2,3\cdots)$.  We call this the ``simple two-slit model (TSS)''.

{\it TSG:} In the second model, in addition to the  the contents of TSS, there is a gas of particles near the slits. These scatter off the electron weakly enough not to disturb its motion but strongly enough to carry away phases.  The gas particles constitute an {\it environment} for the electron  in the sense discussed by \cite{JZ85, Zur03,GH93}  and many others.  We call this the ``two-slit with gas model (TSG)''. 

{\it TSGO:} In the third model the gas  particles in TSG have an observable color --- red near the upper slit and blue near the lower one.  
The electron is replaced by an observer moving through the slits --- so the box must be very large! In these respects  the model is  closer to the real universe where observers are physical systems within the universe and not somehow outside. We call this the ``two-slit with gas and observer model (TSGO)''. Figure \ref{fig2} is a more evocative image of this. 

We stress that in all three examples the boxes  are closed --- like the universe. There are no observers outside the box  looking at the inside, or measuring what goes on there, or otherwise meddling with  the inside.  Any observers are physical systems within a box as in the TSGO model. 

The inputs to the prediction of quantum multiverses are the box's  Hamiltonian $H$ and  its  quantum state $|\Psi(t)\rangle$.  The latter is a function of time in the Schr\"odinger picture in which we work. The state can be defined by specifying the a wave function of the electron at the initial time when it leaves the gun.  This  state evolves to later times by the Schr\"odinger equation as electron moves through the box.  Starting in this way branch state vectors  can be constructed for the individual members of sets of alternative histories of the the electron's motion through the box. The set is decoherent when all the branches are approximately mutually orthogonal. Probabilities for histories are then  the norms of the branch state vectors.  The  computation of these will be described qualitatively  in this section; more mathematical details are described in Appendix \ref{lm}. 
We move back and forth between wave functions like \eqref{wvfn} and bras and kets like $|\Psi(t)\rangle$ as convenient.

\subsection{Simple Two Slit Model (TSS)} Two sets of coarse-grained  histories are readily identified. There is the set of alternative histories defined by which position interval $Y$ the electron arrives at the screen at time $t_d$. A finer grained set of histories is defined by also specifying the slit $S$ that the electron passes through on the way to $Y$.  An even finer-grained set of alternative histories would specify the path the electron takes at each moment of time, but just the  first two sets will be sufficient to illustrate what we need.

Do these two sets of histories define different quantum multiverses that TSS exhibits? Not yet! We still have to check that the sets are decoherent, that is, that the squares of their amplitudes that give probabilities that are consistent with the usual rules of probability theory. 

 A  branch state vector  corresponds to each of the histories in a set of alternative ones. Consider, for example, the set where the histories are labeled only by the different values $Y$ at the screen. The  branch state vectors for this set are
\be
\label{Ybranches}
|\Psi_Y(t_d)\rangle \equiv P_Y |\Psi(t_d)\rangle, \quad Y=1,2,3,\cdots 
\ee
where $P_Y$ is the (Schr\"odinger picture) projection operator onto the range of position corresponding to arrival in $Y$. The collection of projection operators for different $Y$ obey
\be
\label{projs}
P_Y P_{Y'} = \delta_{YY'} P_Y,   \quad  {\sum}_Y P_Y =I 
\ee
capturing the idea that the electron must arrive at one or another of the intervals.  As a consequence, the branch state vectors are orthogonal and sum to the state 
\be
\label{ortho}
\langle\Psi_Y(t_d)|\Psi_{Y'}(t_d)\rangle =0,\  Y\ne Y',  \quad\quad  {\sum}_Y |\Psi_Y\rangle =|\Psi\rangle .
\ee
The orthogonality means that there is vanishing quantum interference between the branch state vectors.  This set of histories is  therefore {\it decoherent}. 
The probabilities $p(Y)$  for arrival at an interval $Y$ are
\be
\label{probY}
p(Y)= || \ |\Psi_Y(t_d)\rangle\ ||^2 = ||\ |P_Y |\Psi(t_d)\rangle\ ||^2. 
\ee

Now, let's consider the finer grained set of histories which specify which slit the electron passed through in addition to where it arrives at the far screen.  The branch state vectors for these histories can be constructed in the same way as in the above example. At the time $t_s$ when the electron passes through the slits define
\be
\label{Sbranches}
|\Psi_S(t_s)\rangle \equiv P_S |\Psi(t_s)\rangle
\ee
where $P_S$ is a projection on a region around slit $S$, either $U$ or $L$. Evolve these vectors to the time $t_d$ and project on the different regions $Y$ to find the branch state vectors 
\be
\label{YSbranches}
|\Psi_{YS}(t_d)\rangle \equiv P_Y |\Psi_S(t_d)\rangle
\ee
for the histories that went through slit $S$ and arrived at interval $Y$.
Evidently
\be
\label{sumamps2}
{\sum}_S |\Psi_{YS}\rangle =|\Psi_Y\rangle.
\ee

Unlike the previous example, this set of histories does  {\it not} decohere. Vanishing overlap does not follow from the orthogonality of the projections expressed in \eqref{projs}. Indeed, the quantum interference between these branches is directly responsible in the classic  two-slit interference pattern.  
  
  It would be inconsistent to try assign probabilities to this set of  alternative histories. The probability to arrive at $Y$ should be the sum of the probability to go through $U$ and arrive at $Y$ and the probability to go through $L$ and arrive at $Y$. But in quantum mechanics probabilities are squares of amplitudes and 
 \be
 \label{inconsist}
||\ |\Psi_Y\rangle||^2 =|| \ |\Psi_{YU}(t_d)\rangle + |\Psi_{YL}(t_d)\rangle \ ||^2  \ne ||  \  | \Psi_{YU}(t_d)\rangle  \ ||^2 +|| \ | \Psi_{YL}(t_d)\rangle \  ||^2
  \ee
because of quantum interference. 

Thus, the finer-grained set following both $Y$ and $S$ does not give us a further quantum multiverse because it is not decoherent. In Copenhagen quantum mechanics we would have said  that there are no probabilities for $S$ because probabilities are assigned only to the outcomes of measurements and  we didn't measure which slit the electron went through.  The Copenhagen approximation is consistent with DH because measured alternatives decohere (e.g. \cite{GH90,Har91a}). But decoherence is a more general, more observer independent criterion for assigning probabilities that apply to histories of the universe as a whole.  

\subsection{ Two-slit Model with Gas (TSG):}

TSG differs from TSS only in the presence of the weakly interacting gas. The weak interaction with the gas means the set of histories coarse grained  only by different values $Y$  will decohere and and have nearly identical probabilities to the same set of histories in TSS. But there will be a significant difference between TSS and TSG for a set of histories that follows $S$ as well as $Y$.  The branch state vectors \eqref{YSbranches} must now include the degrees of freedom of the scattering particles.  Scattering from the upper slit leads to as different state  of the gas than scattering from the lower slit. If enough gas particles scatter these states can be nearly orthogonal leading decoherence of the set of histories that follows both $S$ and $Y$.  This argument is given more quantitatively in Appendix \ref{lm}

Thus for TSG we have  exhibited two multiverses at different levels of coarse graining. The first one (just $Y$) is a coarse graining of the second ($Y,S$). The first ignores the gas, the second follows it. 

Which of these two multiverses  really describes the box model  universe?  Both of them. They are descriptions of the same  system at different levels of coarse graining.  Which of the two descriptions should be used to calculate the probability $p(Y)$ that the electron arrives at interval $Y$ on the far screen?  Either of them because they both supply probabilities that are consistent with one another.

Useful descriptions of a physical system at different levels of coarse graining are familiar from statistical mechanics. A fine-grained  description of a box of gas  would specify the position and momentum of every particle in the box. A much coarser but more useful description specifies only the total energy, angular momentum, and number of particles in the box. These differ greatly in utility and computational complexity.

This model illustrates two of the conclusions in the Introduction.  First (a) even for simple systems like TSG  the theory $(H,\Psi)$ will predict many different multiverses at different levels and kinds of coarse graining.  The two illustrated here are compatible in the sense that one is a coarse graining of the other. But that does not have to be the case. The universe may exhibit incompatible coarse grainings for which there is no finer grained decoherent set of which they are both coarse grainings (think position and momentum.) (See, e.g \cite{GH90}).

The model also illustrates conclusion (b) in the Introduction that multiverses are not a choice or an assumption.  The possible multiverses follow from the the basic theory of the two-slit experiment.  You can choose one multiverse or another to calculate with, but you can't choose  whether the theory exhibits them or not. 

The model also illustrates conclusion (f) in the Introduction. The equivalence of coarse graining by summing quantum amplitudes with coarse graining by summing probabilities is one way of stating the consistency between different sets  that is a consequence of decoherence.  The TSG model gives a specific example. Suppose that we are interested in the probability $p(Y)$ that the electron arrives in interval $Y$. This can be written in two different ways as a consequence of decoherence
\begin{subequations}
\begin{align}
\label{probs-amps}
p(Y)&={\sum}_S\ p(Y,S) = {\sum}_S ||\ |\Psi_{YS}\rangle ||^2 ,   \quad  \text{(sum probabilities)}  \\
p(Y)&= ||\ |\Psi_Y\rangle||^2 = ||{\sum}_S |\Psi_{YS}\rangle||^2   \ \ \quad\quad  \quad \text{(sum amplitudes)} 
\end{align}
\end{subequations}
The first is coarse graining by summing probabilities for histories, the second is coarse graining by summing amplitudes for histories. Which of these formulae should be used to calculate $p(Y)$?  It doesn't make any difference, they both give the same answer  because the probabilities are consistent as a consequence of decoherence. Summing amplitudes is usually easier than summing probabilities because less computation is involved. This is a triviality in this example but, as  we will illustrate in Section \ref{ourbubble},  this simplicity is a considerable advantage in physically complex situations. 

\subsection{Two-slit Model with Gas and Observer (TSGO)}
\label{TSGO}

\subsubsection{Third and First Person Probabilities}
\label{first-third}
TSS and TSG have quantum multiverses that provide probabilities for which of a set of alternative histories happens in the box whether or not they are subject to the attention of observers. These probabilities for which history happens are called {\it third person probabilities} and are derivable just from \HP.   As observers of the universe we are interested in the probabilities for what we observe. To discuss this  TSGO  includes a model observer as a physical system within the box  and  assumes that the gas particles of TSS have a detectable color --- red near the upper slit and blue near the lower slit. See Figure \ref{fig2}.

An observer moves through the two-slit setup  (which must be very large) equipped with a detector that can measure the color of any gas detected. If the detector registers `red' the observer knows that she is  passing through the upper slit. Given this data $red$,  or what is the same thing, given that she passed through  $U$,  what does she predict for the probability that she will arrive in the interval $Y$ at the further screen. This is an example of a  {\it first person probability} --- a probability for the result of an observation \footnote{In other work we have called first person probabilities `bottom up probabilities' and third person probabilities `top-down probabilities'  \cite{HH15c} and especially \cite{HH06}}. First person probabilities  are third  person probabilities for what happens conditioned on  data that describes  observational situation. In this case, the observer's data is that she passed through the upper slit $U$. The first person probability that she observes $Y$ at the further screen is then
\begin{equation}
\label{1stperson}
\pfp(Y) = p(Y|U)=\frac{p(Y,U)}{p(U)} = \frac{||P_Y|\Psi_U(t_d)\rangle||^2}{|| \ |\Psi_U(t_d)\rangle||^2} .
\ee
The distribution is illustrated in bottom image in  Figure \ref{2slit}. As should be clear from comparing that with the figure immediately above  what is most probable to occur is not the most probable to be observed. 

\subsubsection{Anthropic Selection}

Up to now we have tacitly assumed in TSGO  that a live observer (us) exists ($E$) in the box for all times in all cases. Since an observer is a physical system within the box there is a third person probability for it to exist $(E)$ or not exist $(\bar E)$. Our data as observers trivially includes $E$. 

As an example of the consequences of this, of consider histories specified by both $Y$ and $S$. But suppose that the red radiation is lethal to any observer passing through the upper slit.  When we arrive at an interval $Y$ at the far screen we have data $(Y, E)$.  What are the probabilities,  given that data, that we got there\footnote{This is thus a very simple example of  a probability for our past history of the kind that is unavailable in Copenhagen quantum mechanics but which is essential in cosmology as we see in Section \ref{lambda}, See, e.g. \cite{Har98b}} by going through  $U$ or $L$?   Evidently (and trivially) these probabilities are 
\be
\label{anths}
{\pfp}(U) = p(U|Y,E) =0,  \quad \pfp(L) =p(L|Y,E) =1. 
\ee
This is an example of automatic anthropic selection --- point d) in the Introduction.  We wouldn't arrive at $Y$ if we had gone through the lower slit. No `anthropic principle' had to be invoked, no extra ingredient added to \HP. 
We just calculated the  probabilities for what we observe. For more of a discussion of living in a superposition like the observer in this model see \cite{Har15a}. 

\begin{figure}[t]
\includegraphics[width=4in]{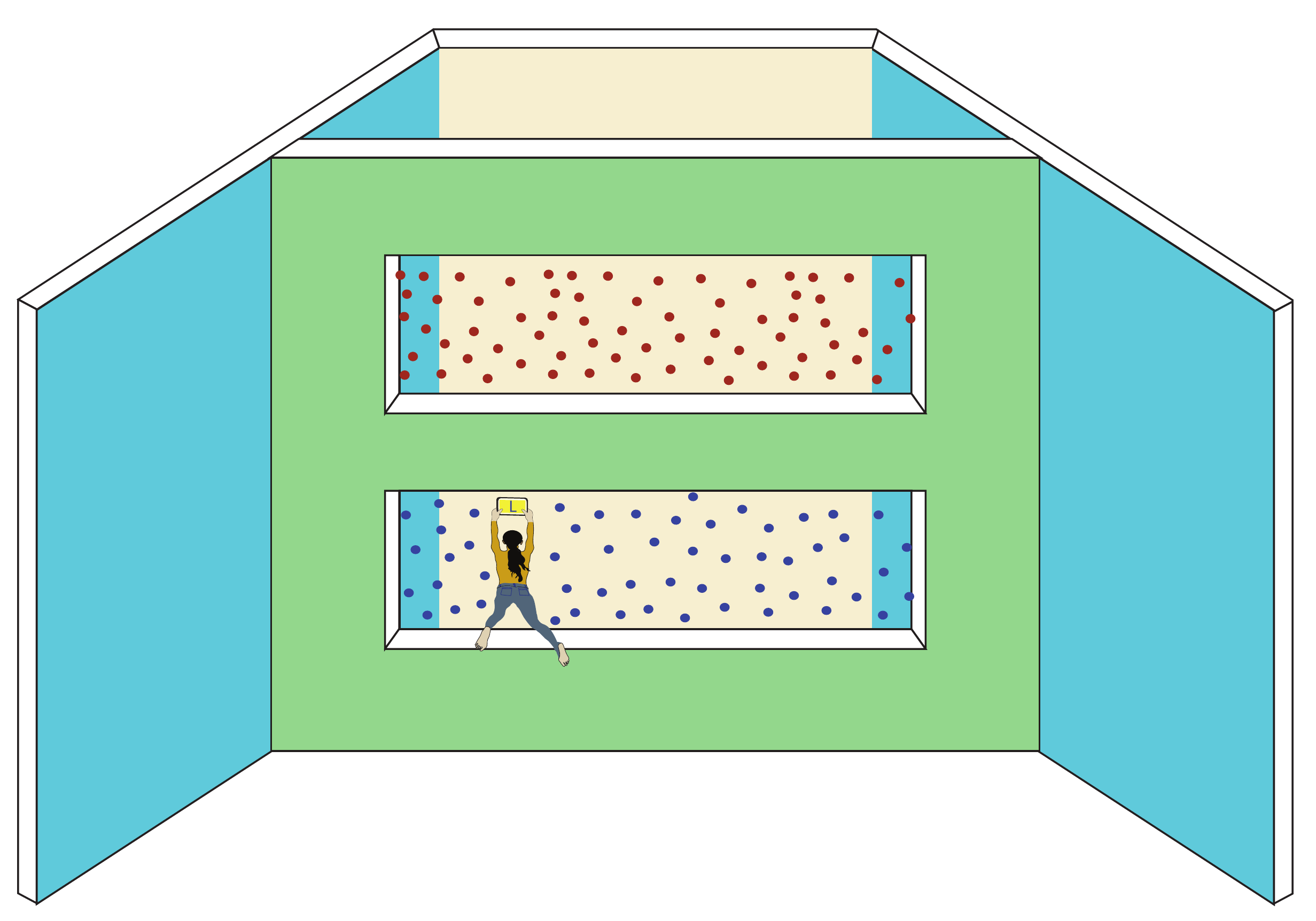}\hfill 
\caption{A more evocative figure of an observer going through a two-slit experiment.  The observer is passing through the lower slit detecting particles of the blue gas which is in her vicinity.}
\label{fig2}
\end{figure}

\section{Five  Exemplary Quantum Cosmological Multiverses}
\label{examples}

The most striking observable feature of our quantum universe is its classical spacetime. At some  level and kind of coarse graining this extends over the whole of the visible universe from near the big bang to the distant future.  What is the origin of this realm of classical predictability in a  quantum theory characterized fundamentally by indeterminacy and distributed probabilities?  How does it emerge from theory \HP\ that includes quantum gravity where spacetime geometry is generally fluctuating and  without definite value?  Many of our cosmological observations are of properties of the universe's classical spacetime and its contents --- the expansion, the approximate homogeneity and isotropy on large scales, the distribution of galaxies, the CMB, the value of the cosmological constant, etc. 
  
Classical behavior is not a given in a quantum universe.  It is a matter of quantum probabilities. A quantum system behaves classically when, in a suitable  quantum multiverse of  alternative histories,  the probabilities are high for those histories exhibiting correlations in time governed by deterministic laws. The relevant  probabilities follow from  \HP. Classical spacetime emerges when the probabilities are high for spacetime geometries correlated in time by the Einstein equation\footnote{The rest of the domain of applicability of classical physics constituting the quasiclassical realm of every day experience is a consequence of this \cite{Har10}.}.

Subsequent sections III through VIII  exhibit five models of a quantum multiverse of classical spacetimes that illustrate conclusions  (a)-(f) of the Introduction.  The models are based on the author's joint work with Stephen Hawking and Thomas Hertog (mainly \cite{HH15b,HH13,HHH10b,Her14}).  We do not pretend to present these models in the depth and  precision that they are discussed in those papers.  Rather, we present a qualitative and intuitive  descriptions of the models.

\subsection{Common Elements}
\label{common}

The five models have the following elements in common:

\subsubsection{General Theory}
\label{genth}

{\it Semiclassical Quantum Gravity:|} Quantum multiverses of the universe necessarily include alternative histories of cosmological spacetime geometry and therefore involve quantum gravity at some approximate level. Phenomena like the emergence of classical spacetime from the quantum big bang, the generation of large scale structure from quantum fluctuations away from homogeneity and isotropy,  the nucleation of bubbles of true vacuum in a false vacuum, and eternal inflation are all fundamentally quantum spacetime phenomena. We discuss these treating  quantum gravity in semiclassical approximation.

{\it Quantum State:} For the quantum state  $\Psi$ we adopt the no-boundary wave function of the universe (NBWF) in its semiclassical approximation \cite{HH83}. 

{\it Dynamics}: For  a theory of dynamics $(H)$ we assume Einstein gravity coupled to a single homogeneous scalar field $\phi(t)$ moving in a potential $V(\phi)$. Different models have different $V(\phi)$

{\it Coarse Graining:} The  theory \HP\ predicts different quantum multiverses defined by coarse grainings that follow different variables on very  different scales.  Coarse grainings relevant for laboratory experiment follow the outcomes of the experiment and ignore cosmological scale features of the universe if these do not affect the outcomes. Coarse grainings relevant for cosmology follow the large scale features of the universe and ignore small scale fluctuations like planets, biota, human observers, and their laboratory experiments whose presence or absence has little effect on the large scale behavior of the universe. 

{\it A Model of Observers and Observation:} 
 \label{firstperson}
 By itself,  the theory \HPs predicts third person probabilities for which of a set of alternative histories of  classical spacetime geometries and the matter within happens.    First person probabilities for our observations are third person probabilities for what happened conditioned on the data $D$ describing our observational situation.  TSGO provided a very simple example of this leading to \eqref{1stperson} and\eqref{anths}. There, the data $D$ was that an observer existed, $E$. Naturally probabilities for the results of her observations were conditioned in $E$.  

As observers of the universe we, and the apparatus we use, are quantum physical systems within it not somehow outside it. Our observations of the universe are limited to a spatial volume of rough size $c/(\text{Hubble constant})\sim 4000Mpc$ at the present time  --- our Hubble volume.  This  is just one Hubble volume in a universe that may have a great many similar volumes. We have only a very, very small probability  that we denote by $p_E(D)$ to have evolved in any Hubble volume. Incorporating, as it does, the probabilities for accidents of biological evolution, the probability $p_E(D)$ is a very, very, very small number  much beyond the ability of present day physics to compute. This is a very crude model of an observing situation, but still better than many treatments where the evolution of observing systems in the universe is not considered at all.\footnote{For detail on observers, observations 1st and 3rd person probabilities and what's included in $D$  etc see  e.g.\cite{HS07,HH15c}.}

Despite the small value of $p_E(D)$ there can be a significant probability that in a very large universe the data $D$ is replicated in many Hubbble volumes. 
First person probabilities for our observations are third person probabilities for what happened conditioned on the existence of at least one instance of our observational situation --- our instance.  That's all we know for sure about instances of $D$. We abbreviate (at least one instance of $D$) by $\Dge$. TSGO provided a very simple example of this leading to \eqref{1stperson} and\eqref{anths}. 

The bottom line is that first person probabilities for an observable feature of the universe $\cal O$ are third person probabilities for $\cal O$ to happen from \HP\  conditioned  on the existence of at least one instance of our observational situation $D$, viz
\be
\label{probobs}
p^{(1p)}({\cal O}) \equiv p({\cal O}|D^{\ge 1})
\ee

We describe how to calculate this in the various models beginning with the one in the next section. 

\section{ A Quantum Multiverse of Homogeneous, Isotropi, Classical Histories}
\label{homoiso}

This is a very simple model that illustrates how a quantum multiverse of classical spacetime geometries emerges from \HP.

The model assumes  a minisuperspace based on    homogeneous, isotropic, spatially  closed spacetime geometries with metrics of the form
\be
ds^2 = -dt^2 +a^2(t) d\Omega_3^2 . 
\label{metric}
\ee
Here, $a(t)$ is the scale factor and $d\Omega_3^2$ is the metric on a unit round 3-sphere.
 
 {\it Dynamics}: For  a theory of dynamics $(H)$ we assume Einstein gravity and a single homogeneous scalar field $\phi(t)$ moving in a potential $V(\phi)$. with the form:
 \be
\label{simplepot}
V(\phi)= \Lambda + \frac{1}{2} m^2\phi^2
\ee
in the Planck units $(\hbar=G=c=1)$ that we use throughout. 
 An action  $I[(a(t),\phi(t)]$ summarizes this dynamics but we won't need its explicit form.

 {\it Quantum State:} The NBWF is  a function(al) of the geometry and field configurations on a spacelike three-surface. In this minisuperspace model this means that the wave function depends on the scale factor of the three surface and the homogeneous value of the scalar field  there. When used as arguments of the wave function we denote these by $b$ and $\chi$ respectively. Thus, 
\be
\Psi=\Psi(b,\chi).
\label{wf}
\ee
In the semiclassical approximation the NBWF is a sum of terms of the form
\be
\Psi(b,\chi) \propto \exp[-I(b,\chi)/ \hbar]
\label{nbwf}
\ee
where $I(b,\chi)$  is the action at a saddle point of  the action $I[(a(t),\phi(t)]$ that  is regular on a four-disk and for which $(a(t),\phi(t))$ assume the values $(b,\chi)$ on its  boundary. These saddle points are generally complex so we can write
\be
\Psi(b,\chi) \propto \exp\{-[I_R(b,\chi)+iS(b,\chi)]/\hbar\} 
\label{ri}
\ee
where $I_R$ and $-S$  are the real and imaginary parts respectively of the saddle point action. 

  {In regions of $(b,\chi)$  where $S$ varies rapidly compared to $I_R$ this wave function takes a WKB form and predicts a multiverse of 
Lorentzian, classical histories obeying the Einstein equation \cite{rules}.  This classical multiverse  consists of the the integral curves of $S$. That is, the  histories are  the solutions of  the Hamilton-Jacobi expressions for the momenta $\pi_b$ and $\pi_\chi$ conjugate to $b$ and $\chi$, viz.  
\be
\label{intcurves}
\pi_b  = \frac{\partial S}{\partial b}, \quad \pi_\chi = \frac{\partial S}{\partial \chi}  .
\ee

There turns out to be  a  one parameter family of such curves conveniently labeled by the magnitude of the scalar field  $\phi_0$ at the center of the saddle point geometry.   The label $\p0$ turns out to be approximately equal to the initial value at which the field starts rolling down the potential in the classical history labeled by $\p0$. The probabilities of these histories are approximately 
\be
p(\phi_0) \propto \exp[-2I_R(\phi_0)/\hbar] ,
\label{probhist}
\ee
which, for this model, is approximately
\be
\label{3p}
p(\phi_0) \propto  \exp\left[\frac{3\pi}{\Lambda+(1/2)m^2\phi_0^2}\right] .
\ee

Thus, the theory \HPs predicts a quantum multiverse of classical spacetime geometries  and matter fields labeled by $\phi_0$ with probabilities \eqref{3p}. We can say that these are the third person  probabilities that the history labeled by $\p0$ was the history of the universe that happened or occurred.

This multiverse does not provide a mechanism for the variation of $\Lambda$. There is only one minimum characterized by one value of $\Lambda$ and all the possible histories roll down to that. The same is true for $m$ which turns out to govern the size of the primordial density fluctuations observed in  the CMB  \cite{HHH10a}. 

\subsubsection{Probabilities for Inflation}
\label{obsobs}
The number of e-folds of scalar field driven inflation $N_e(\p0)$  is a simple example of a quantity which, if not directly observable, has significant observable consequences for our universe. Numerical solution of the equations \eqref{intcurves} for the individual members of the multiverse of classical histories shows that approximately \cite{HHH08}.
\be
\label{nefolds}
N_e(\p0) \approx  3\p0^2/2
\ee
for $\p0 \gtrsim1$  in the Planck units used throughout. 
The third person probabilities for \eqref{3p} show that histories with a low amount of inflation are the most probable to occur. But what is most probable to occur is not necessarily the most probable for us to observe.  Evaluating \eqref{probobs} gives those probabilities. 

From the definition of conditional probabilities (Bayes theorem) we have 
\be
\label{bayes} 
p(\p0|\Dge) \propto p(\Dge|\p0)p(\p0) .
\ee
Normalizing the right hand side gives a formula with an equality.  In this simple model of the observing situation 3rd person probabilities for what occurs are converted to 1st person probabilities for what is observed by multiplying by the factor $p(\Dge|\p0)$ called the `top-down' factor and then renormalizing.

The probability that there is at least one instance of our observational situation somewhere in the universe, $p(\Dge|\p0)$  is bigger in a larger universe where there are more Hubble volumes  for an  instance of $D$  to have evolved than it is  in a smaller universe where there are fewer Hubble volumes.  This turns out to mean that we are more likely to observe universes with more inflation \cite{HHHsum}.  

We can understand this result more quantitatively with a little model where $p(\Dge|\p0)$ can be explicitly evaluated. 
Suppose that our data $D$ locate us somewhere on a homogeneous isotropic spacelike surface with $N_h (\p0)$  Hubble volumes in each classical history.  The probability $p(\Dge|\p0)$ is one minus the probability that there are no instances of $D$ on the surface. This in turn is the product of the probabilities $1-p_E(D)$ that there are no instances of $D$ in any particular Hubble volume on the surface. The result is the formula:
\be
\label{tdfactor}
p(\Dge|\p0)=1-[1-p_E(D)]^{N_h(\p0)}. 
\ee
Eq. \eqref{tdfactor} is an explicit formula for the `top-down factor' that converts 3rd person probabilities to first person ones  as in \eqref{bayes}.  Larger number of Hubble volumes $N_h$ make this factor larger and closer and closer to 1.  However small $p_E(D)$ is, in a sufficiently large universe the probability  is $1$ that an instance of $D$ occurs somewhere. When $N_h\gg1/p_E(D)$ first person probabilities are equal to third person ones.  

First person probabilities thus favor larger universes, larger $\p0$, with more e-folds for inflation \eqref{nefolds}.
In the quantum multiverse considered in this section significant inflation is anthropically selected to be observed.

\subsubsection{How this Model Supports the  Conclusions}
\label{cont1a}
The main contribution of this model to our exposition is to illustrate in a simple way how quantum multiverses of classical histories (including spacetime geometry) are predicted by \HP. We will assume this for subsequent models.  But the model does illustrate points (b) and (c) of the Introduction, namely that quantum multiverses are a consequence of \HP\ and typically consist of many different histories not just one.

\begin{figure}[t]
\includegraphics[width=6in]{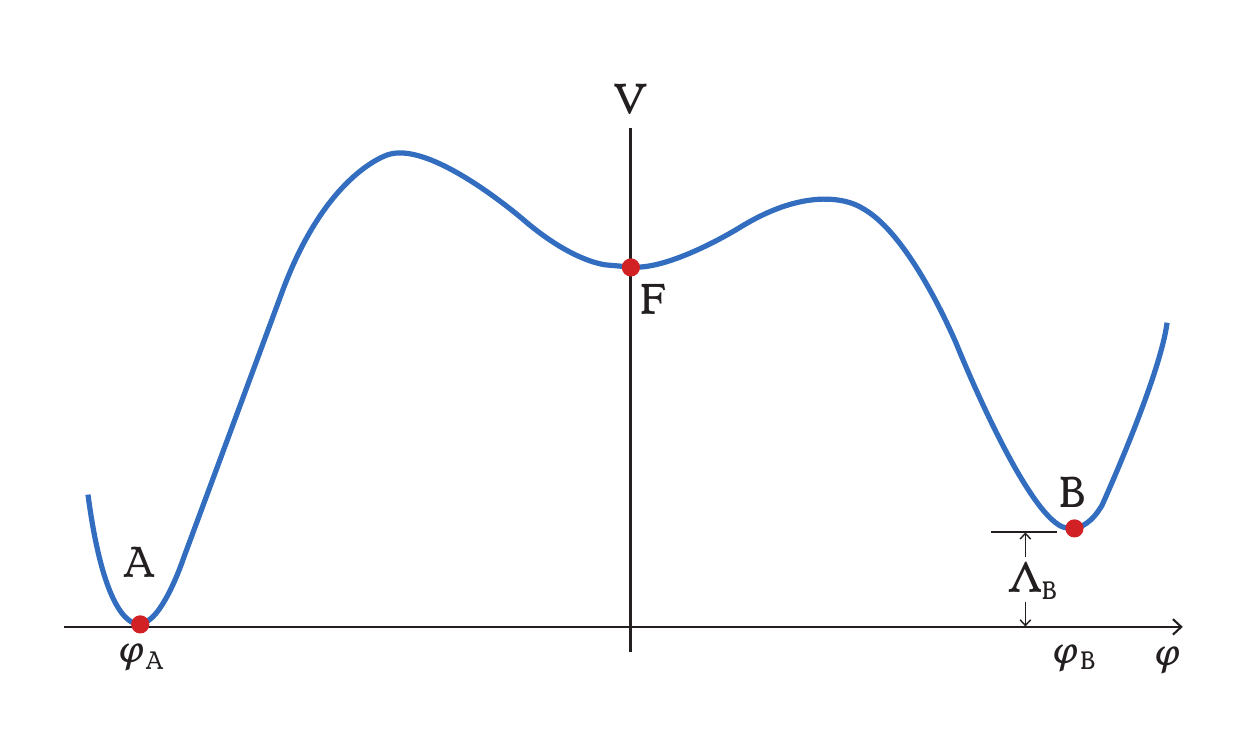}\hfill 
\caption{A potential for the scalar field with one false vacuum $F$ and two true vacua $A$ and $B$. The false vacuum is separated from either true vacua by potential barriers and  relatively flat patches where the conditions for slow roll inflation are satisfied. The different shape of the barriers and of the potential in the two slow roll regimes leading to the true vacua results in different false vacuum decay rates and different predictions for CMB related observables in universes ending up in $A$ or $B$. }
\label{pot}
\end{figure}

\section{ Quantum Multiverses of Pocket Universes}
\label{bubbles}

This section shows how pocket multiverses mentioned in  the Introduction arise at various levels of coarse graining from the NBWF $(\Psi)$ and a dynamical theory $(H)$ based on a particular potential $V(\phi)$  like the one in Figure \ref{pot}. This potential has  three minima (vacua) --- two true vacua $A$ and $B$ and one false vacuum $F$.  We can say that the potential defines a {\it landscape of vacua} although there are only three here. As described in Section \ref{homoiso},   \HPs predicts a one parameter ensemble of classical histories labeled by the value  at which they start to roll down.  

The classical history that starts to roll down at  $F $   in Figure \ref{pot} is a universe that inflates with an effective cosmological constant $3\pi/V(0)$.
Classically this inflation is eternal --- goes on forever. But quantum mechanically regions of spacetime can tunnel through the barriers on either side of $F$ and roll down classically to one of the true vacua $A$ or $B$. That is, the inflating background can nucleate bubbles of true vacuum $A$ or $B$. The result is a multiverse whose Penrose diagram is like the one in Fig \ref{fine-grained} for low nucleation rates. Different histories will have different numbers of true vacua of either kind in different places in the eternally inflating universe. 

\begin{figure}[t]
\includegraphics[width=4in]{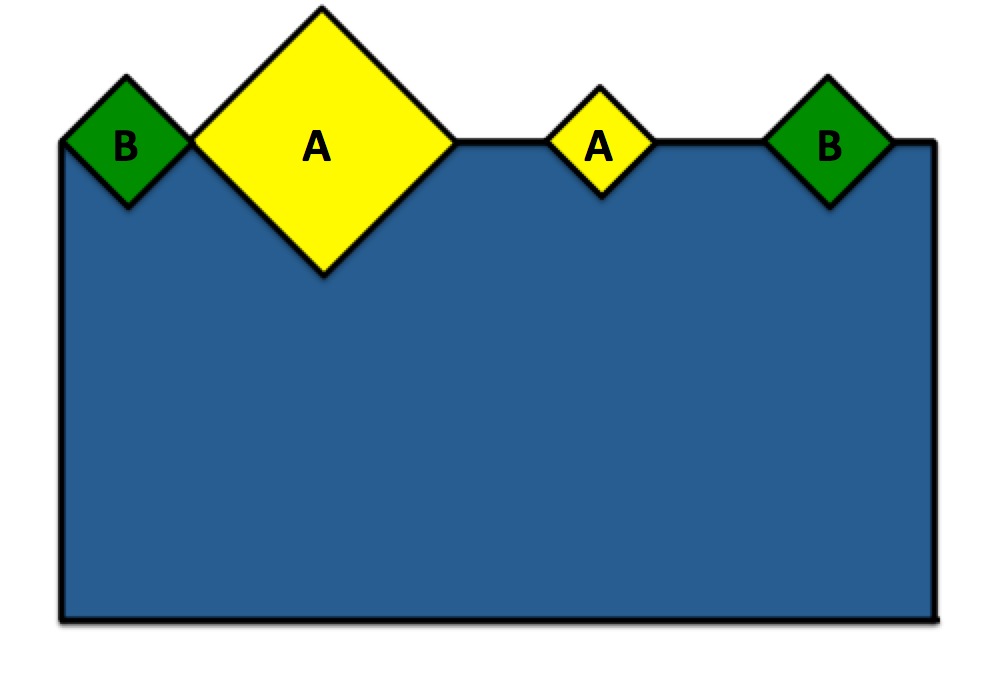}\hfill 
\caption{The Penrose Diagram for the pocket multiverse model when the bubble nucleation rate is low enough that  collisions between bubbles are negligible. The blue (dark) region in the diagram represents the false vacuum  classical deSitter expansion. That deterministic classical evolution  is interrupted by the  nucleation of  bubbles of true vacuum in the false vacuum by quantum tunneling which then expand at the speed of light.  The bubbles  nucleate at a point in the false vacuum and their walls expand at the speed of light. There are two kinds of true vacuum $A$ (yellow) and $B$ (green) corresponding to the two minima in the potential in Figure \ref{pot}. }
\label{fine-grained}
\end{figure}

We don't live in the false vacuum devoid of matter. Rather we live in one of the bubbles of true vacuum, either one of type $A$ or of type $B$. The kind of bubble we live in can be determined from observation. The value of the potential at the the minimum determines the local value of the cosmological constant, either $\Lambda_A=V(\phi_A)$ or  $\Lambda_B=V(\phi_B)$. Observations of the expansion can in principle determine the cosmological constant in our bubble and therefore determine which kind of bubble we are in. The curvatures of the potential at the minima are related to the the amplitude of primordial density fluctuations which leave an imprint on the $CMB$  \cite{HHH10a}.  Different curvatures in at the minima of $V$ produce observationally distinct $CMB_A$  and $CMB_B$. By observing  the CMB which we can determine which kind bubble we are in. 

This model supports several of the conclusions in the Introduction.   It supplies  a concrete example of how certain \HPs  implies a pocket multiverse that is realized as a false vacuum eternally inflating universe that nucleates bubbles of true vacuum (the pockets) with different observational productions, In particular for a suitable potential it shows how the cosmological constant can vary from bubble to bubble in space. This pocket universe is not a choice nor postulate once an appropriate \HP\ is fixed. Were $\Lambda_B$ too large to form galaxies by as described in the Introduction we would immediately predict we are living in a bubble of type $A$, That is a simple example of anthropic selection but we will have more to say about this in the next section.

\begin{figure}[t]
\includegraphics[width=4in]{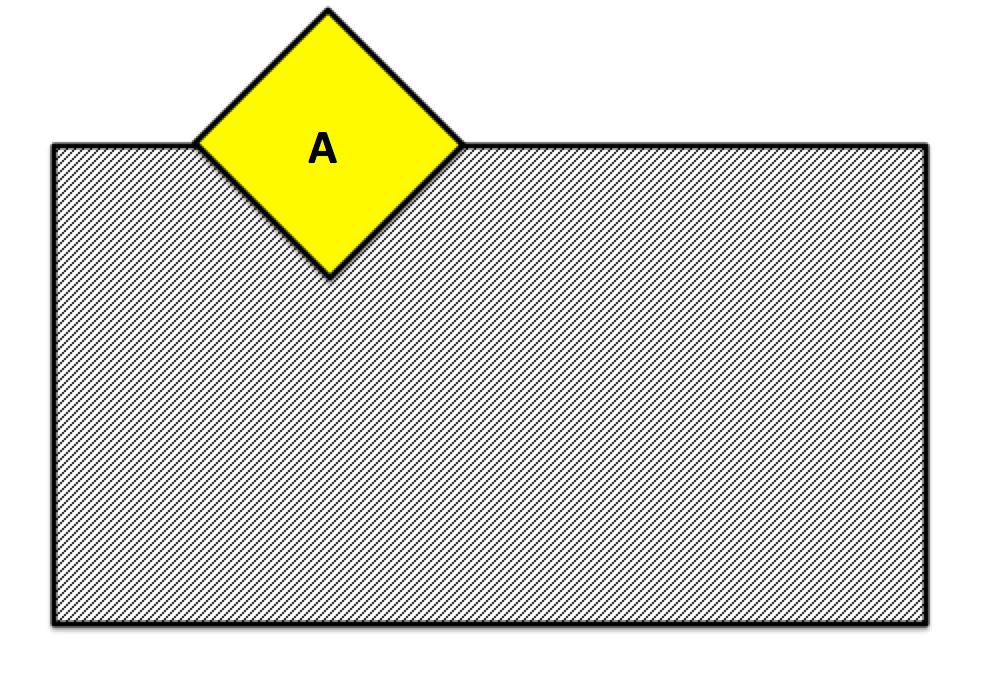}\hfill
\caption{A history of the universe that is much coarser grained than the one in Figure \ref{fine-grained}. There is only one bubble --- our bubble.  The mosaic of false vacuum $A$ and $B$ bubbles that are outside our bubble  suggested in Figure \ref{fine-grained} is ignored and shaded gray.  We stress that this does not mean there are no regions of $A$, $B$, or $F$ outside. Only that the coarse graining does not distinguish whether a given region outside our bubble is $A$, $B$, or $F$. }
\label{cgbubbles}
\end{figure}

\section{A Quantum Multiverse of the Observable Properties of Our Bubble}
\label{ourbubble}
In the TSG model in Section \ref{dh}  we showed how one quantum system can exhibit different quantum multiverses at different levels of coarse graining. In this section we show how the model cosmology defined by the potential  like that in Figure \ref{pot} can exhibit more than one quantum multiverse at different levels of coarse graining. 

A relatively fine-grained multiverse  would consist of histories that describe all possible configurations of bubble nucleation and non-nucleation at all places and all times.  The history illustrated in Figure \ref{fine-grained}  is just one example. But there are an infinite number of other histories at this level of coarse graining in a false vacuum deSitter expansion that extends to the infinite future. 

The (3rd person) probabilities per unit four volume  $p_A$ and $p_B$  for nucleating bubbles of different kinds  were calculated by 
Coleman and De Lucia (CDL) \cite{CDL}. The probability per unit four-volume to  remain in the false vacuum is $p_F=1-p_A-p_B$. 
With these probabilities  it is possible to imagine calculating the third person probabilities for an entire set of histories of different mosaics of volumes of false and true vacua. Some kind of cutoff or `measure' would be required to deal with the infinite volume. Even then it would be a formidable calculation. 

We do not observe entire four-dimensional histories of roiling seas of bubble nucleation extending to the infinite future. Neither do we know the location of our bubble. First person probabilities for our observations, say of the CMB, are the probabilities that we are in a bubble of type $A$ or $B$ at some unknown location in the eternally inflating false vacuum spacetime. Such probabilities could in principle be calculated by summing the probabilities for the fine-grained histories over the all the alternative structures outside our bubble.  Quantum mechanics provides a more direct route to the answer:  Make predictions for our observations using a quantum multiverse based on a  much coarser grained multiverse that follows what goes on inside our bubble and ignores everything outside --- a multiverse based on our bubble.  

A history in  our bubble multiverse is illustrated in Figure \ref{cgbubbles}. Our bubble occupies one part of the spacetime with (CDL) probabilities $p_A$ or $p_B$ that it is of type $A$ or $B$. Since the coarse-graining doesn't specify what is going on in the volumes outside our bubble their contribution to the probability of this coarse-grained history is $p_A+p_B+p_F=1$.  

Suppose the data $D$ describing our observational situation locate us on the spacelike reheating surface inside one of these kinds of bubbles. There are an infinite number of Hubble volumes on these surfaces in both kinds of bubble. The top-down factor \eqref{tdfactor} connecting 1st and 3rd person probabilities is then unity.  1st person probabilities equal 3rd person probabilities. Thus,
the probabilities that we observe $A$, $p(WOA)$ or $B$ $p((WOB)$ are 
\be
\label{obs}
p(WOA) = \frac{p_A}{p_A+p_B}, \quad p(WOB) = \frac{p_B}{p_A+p_B}
\ee
It doesn't matter where in spacetime our bubble is located. CDL probabilities respect the symmetries of deSitter space and are the same at all locations. 

This example illustrates that a quantum system like our universe is not described by only one quantum multiverse. It is described by many different ones at different levels of coarse graining within the same theory \HP\ with the same potential in Figure \ref{pot}.  
For a given question it's generally best to use the coarsest grained description that supplies an answer  to it.

This model supports all of the conclusions (a)-(f) mentioned in the Introduction (a) Different quantum multiverses follow from \HP\ at different levels and kinds of coarse graining. (b) They emerged by calculation from \HP with the potential in Fig \ref{pot}. They were not an assumption beyond assuming \HP.  (c.) The coarse graining following the alternatives $A$, $B$, or $F$ for all volumes in spacetime leads to quantum multiverse with a truly vast number of histories to compute third person probabilities for. The coarse graining following only the inside of our bubble and ignoring everything outside has only  alternatives $A$ or $B$. Either way there is a multiplicity of alternative histories. (d.)  Anthropic selection automatically ruled out an observation of a false vacuum $F$ devoid of matter where we cannot exist. (f.) The simplicity of coarse graining  multiverses  by summing amplitudes  rather than summing probabilities.

\subsection{Different Dynamics, Different Multiverses}
\label{different}
Our story about pocket universes forming by the decay of an eternally inflating false vacuum depended crucially on a dynamical theory  $(H)$ incorporating something like the potential  in Figure \ref{pot}. This dependence on dynamical theory is illustrated in work by Hawking and Hertog \cite{HH17}.  They use the no-boundary quantum state $(\Psi)$ together with a dynamics $(H)$ specified by an effective dual field theory defined on the exit surface of eternal inflation. This dual field theory provides a description of the transition between the essentially quantum realm of eternal inflation to the ensemble of possible classical universes one of which we observe. It provides a quantum probabilistic measure on this ensemble that is different from the one in Section \ref{ourbubble}.  Hawking and Hertog find that this measure predicts a smooth exit from eternal inflation with no bubbles and no pocket universes. Evidently this dynamics is not represented by a an effective dynamical theory incorporating a potential like that in Figure \ref{pot}.

\section{A Quantum Multiverse of Homogeneous and Isotropic Classical Histories with Different  Physical Constants}
\label{lambda}. 

\begin{figure}[t]
\includegraphics[width=6in]{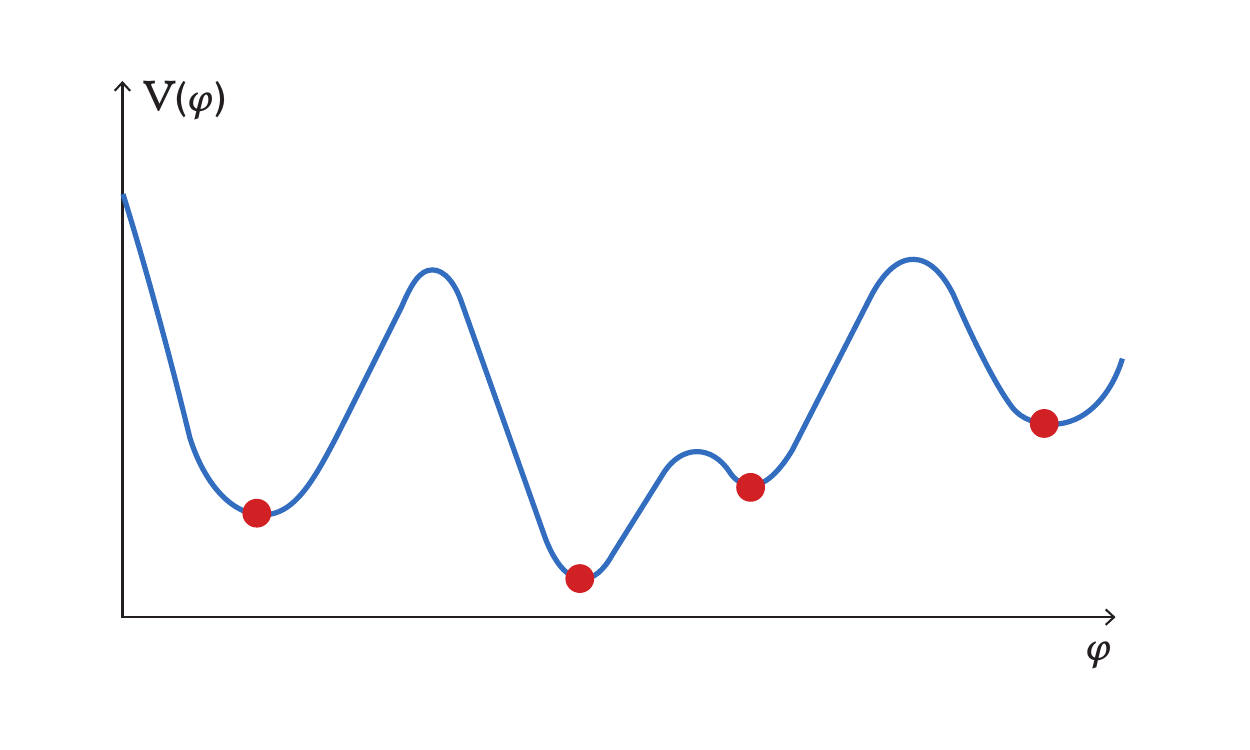}
\caption{A potential $V(\phi)$ for the scalar field with many minima at field values $\phi_K, K=1,2,\cdots$.  The values of the field at the minima define different values of the cosmological constant $\Lambda_K=V(\phi_K)$.  The first person probability that we observe one value $\Lambda$ or another  is the probability that our past history rolled down to one minimum or another. }
\label{landscape}
\end{figure}
The potential on which this model is based is  shown in Figure \ref{landscape}.  It has many minima (vacua)  $K=1,2,\cdots$  at values $\phi_K$ near which it is approximately 
\be
V(\phi) \approx   \Lambda_K + \frac{1}{2} m_K^2  (\phi-\phi_K)^2 +\cdots
\label{minima}
\ee
for constants $\Lambda_K$ and $m_K$ --- defining a landscape of vacuua.

This potential defines a one parameter  ensemble of  alternative homogenous and isotropic classical histories that may roll down to one minimum or another. The theory \HPs supplies  3rd person probabilities for which minimum is reached.  The ensemble is thus a quantum multiverse in which the constants $\Lambda$ and $m$ are the same everywhere in space but vary from history to history.  Their value is thus not fixed by the action but by the quantum accident of which one of the minima is reached. 
\begin{figure}[t]
\includegraphics[width=4in]{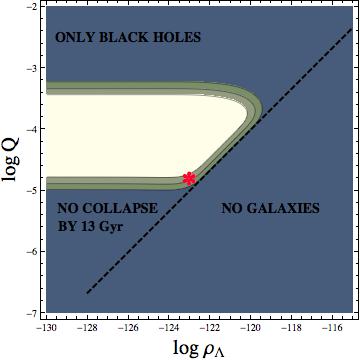}
\caption{Anthropic  constraints on the values of $Q$ and $\Lambda$ in a figure from \cite{HH13} based on calculations in Tegmark, et. al.  \cite{Tegetal06} given the present age of the universe.  }
\label{lam-landscape}
\end{figure}

As observers of the universe we are a special kind of late time fluctuation away from homogeneity and isotropy living
in  one of the true vacua after reheating has produced ordinary matter. The values of $\Lambda$ and $m$  that we observe are the values for our minima. The parameter $m$ is related to the amplitude $Q$  of primordial density fluctuations when they leave their horizon.  The relation turns out to be roughly $Q\approx m$ in the Planck units employed throughout. The 1st person probability that we observe particular values of $\Lambda$ and $Q$ is a probability that our past history rolled down to a minimum with these values\footnote{The  approximate Copenhagen quantum mechanics of measured subsystems would not be able to give us these probabilities because it cannot generally retrodict the past  (e.g \cite{Har98b}). A generalization like DH is necessary to do that.}.

In Section \ref{obsobs} we described how the transition between 3rd person probabilities for which history occurs and 1st person probabilities for which history is observed favors larger universes where there are more places of us to be and disfavors smaller universes. The history that dominates the 1st  person probabilities  for observation turns out to be the one with the lowest $\phi_0$ consistent with significant inflation. In more technical terms 1st person probabilities are dominated by the  lowest exit from eternal inflation. As explained in Section \ref{obsobs} for large universes with many Hubble volumes the top-down factor \eqref{tdfactor} is unity  and  1st person and 3rd person probabilities are equal.

  As mentioned before it turns out  that $m\approx Q$ so  the 1st person probabilities for $\Lambda$ and $Q$  then are approximately  from \eqref{3p}
\be
\label{1st-person}
p^{(1p)}(\Lambda, Q) \approx  \exp\left(\frac{3\pi}{\Lambda+ c Q}\right) .
\ee
where $c$ is a calculable dimensionless constant of order unity. 

\begin{figure}[t]
\includegraphics[width=6in]{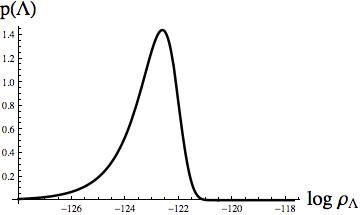}
\caption{The marginal probability distribution for observations of the the cosmological constant obtained essentially by integrating the 1st person probabilities \eqref{1st-person} over $Q$}. 
\label{NBWF-Lambda}
\end{figure}

Conditioning on at least one instance of a small amount of our data on the scales of the Earth turns out to be  sufficient to derive \eqref{1st-person}. We can then use this result  to look for predictions of correlations between the rest of our data on scales of our Hubble volume. To illustrate this we look for correlations among three pieces of our data:  the value of $\Lambda$, the value of $Q$ , and the fact that we have a Hubble volume  full of galaxies by the present age of approximately $14$Gyr.  The range of values of $\Lambda$ and $Q$ for which these conditions can satisfied as calculated in \cite{Tegetal06} and displayed in 
Figure \ref{lam-landscape}.   As noted by \cite{BT86,Wei89}, too large a $\Lambda$ would prevent proto-galaxies from collapsing. Too large a $Q$ would result in a universe where most collapses produce black holes.  Were $Q$  too small the fluctuations would not grow enough for galaxies to form by the present age. These anthropic constraints mean that the probability is negligible that we are outside  the white region in Figure \ref{lam-landscape}.  

Inside the white region the NBWF 1st person probabilities \eqref{1st-person} favor small values of $Q$ and larger values of ${\rm log}_{10}\Lambda$. Thus we predict the values indicated by the red $\color{red}\bigstar$
\be
Q\sim 10^{-5}, \quad\quad \Lambda \sim 10^{-123} 
\label{pred}
\ee
Figure \ref{NBWF-Lambda}  shows the  marginal distribution for $\Lambda$.

The few orders of magnitude agreement of the values in \eqref{pred} with observations is not the main point. After all this is just one example out of many possible ones. Rather the main things the reader should take away is the following: First, the universe does not have to have pockets for the fundamental constants to vary. As here,  there can be quantum multiverses of homogeneous and isotropic histories in which the constants do not vary in space but over a multiverse of histories. Second, anthropic selection is automatic in quantum cosmology through 1st person probabilities for observations. And finally, a theory of the quantum state has a significant impact on what the  probabilities are and for our predictions of the 1st person probabilities of the constants we observe. 

\section{A Quantum Multiverse of Histories with Different CMBs}
\label{st}

In the preceding section's calculation of 1st person probabilities for observation only the potential in Figure \ref{pot} below the lowest exits  from eternal inflation contributed to the final result.  The potential is thus effectively equivalent to an ensemble of one dimensional potentials each with one minimum. 

String theory has a vast landscape of possible vacua \cite{DD04}.  A landscape with  many fields moving in an ensemble of one-dimensional potentials each with one minimum is probably more analogous to  the string landscape than one potential with many minima of the kind in Figure \ref{pot}. Thomas Hertog used such a model landscape consisting of a number of one dimensional potentials for $(H)$ together with the no-boundary wave function  for $(\Psi)$ to  estimate that theory's  prediction for the  tensor/scalar ratio in the CMB \cite{Her14}.   For a model landscape he took the scalar field potentials that were used to reduce data from the Planck satellite on the CMB  \cite{Planck}. That landscape included power law potentials of the form $V(\phi)=\lambda \phi^n$, plateau potentials of the form $V(\phi) =V_0(1-\phi^n/\mu)$ and  `$R^2$ inflation potentials'.  The quantum multiverse consists of all the histories in all the potentials. The most probable 1st person history is the one with the lowest exit from eternal inflation among all these potentials.

Hertog's results are shown in Figure \ref{prior}.  The important point about the model is not the specific numerical value for the tensor/scalar  ratio  but the fact that they are a prediction for observations yet to be done.  If further observations yield different numbers then either the NBWF is ruled out or the ad hoc landscape of potentials would be ruled out.

\begin{figure}[t]
\includegraphics[width=6in]{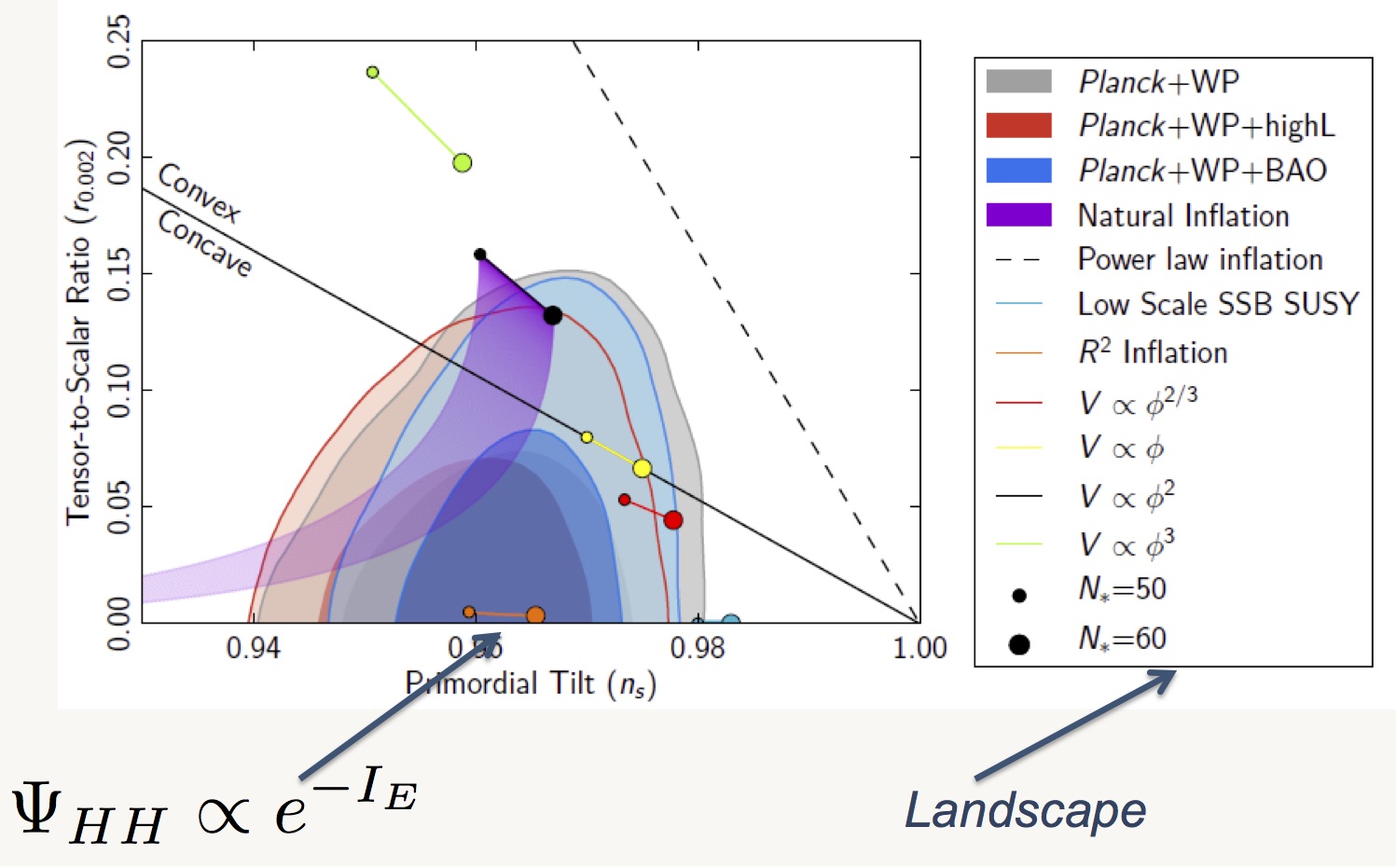}\hfill 
\caption{Planck satellite results for  the tensor/scalar ratio $r$ and the scalar spectral tilt $n_s$ of the observed CMB as compared with various theories.  The figure was adapted by Thomas Hertog from Figure 1 of \cite{Planck}. The shaded regions show the values consistent with the data when combined with other astronomical measurements. The predicted curves are predictions from various kinds of models.  The NBWF plus the landscape of potentials discussed in the text and labeled at right predicts the values indicated roughly by the arrow. }
\label{prior}
\end{figure}

\section{Conclusions}
\label{concl}

At the start of this paper we defined a multiverse as an ensemble of alternative possible situations only one of which is observed by us. We have seen how  decoherent histories quantum mechanics (DH)  predicts quantum cosmological  multiverses in the form of decoherent sets of alternative coarse grained histories of the universe. In each set one history occurs, or happens, with a probability predicted by a fundamental quantum theory of cosmology consisting of  theories of the universe's  quantum state ($\Psi$) and dynamics ($H$).   Quantum multiverses are not some posited speculative idea to be grafted onto the basic theory. They are the output  of that theory.

\subsection{General Conclusions}
\label{genconcl}

Conclusions that follow from  general perspective of this paper  have been stated at several points in this paper but we repeat them here in abbreviated form\footnote{Many years ago John Wheeler explained to the young author  that some people only read  the introduction, some people only read the conclusions, and some people only read the figure captions, so your message should be in all of those places.}.

{\it a. Many Quantum Multiverses.} The theory \HP\ does not just predict one multiverse of alternative histories. It predicts many different multiverses at different levels and kinds of coarse graining.  The pocket multiverse is just one example.

{\it b. Multiverses are Not a Choice.} Quantum multiverses are not a choice or an assumption separate from the theory \HP. They follow or do not follow from \HP.  If you have some prior objection to multiverses of some kind then restrict to theories \HP\ that don't imply them.

{\it c. Multiverses are Generic.} Simple, manageable, discoverable  \HP's generically predict quantum multiverses consisting of an ensemble of  many possible histories together with probabilities for which one occurs.  If the ensemble consisted of  just one history with  certainty  all of present complexity would have to be encoded \HP. 

{\it d. Anthropic Selection is Automatic.} Anthropic selection is an automatic consequence of  first person quantum mechanical probabilities for observations (Section \ref{first-third}). Probabilities for our observations are conditioned on a description of our observational situation and we won't observe what is where we cannot exist. 

{\it e.  Different Mechanisms for the Variation of Constants.}  In different quantum multiverses constants like $\Lambda$ can be constant in each history but very from history to history, or can vary in space within every history.  Automatic anthropic selection is enabled  in either case. 

{\it f. Two Routes to Coarse-Graining.}  Quantum multiverses are restricted to sets of alternative coarse grained histories that are decoherent --- that have negligible quantum interference between individual histories in a set. Further coarse graining can be carried out either by summing probabilities or by summing quantum amplitudes. That is a considerable computational advantage. 

Understanding quantum multiverses as decoherent sets of alternative coarse-grained histories of the universe provides a unified perspective on different notions of multiverse that have been discussed. As in the two-slit models in Section \ref{dh} it includes the multiverses that describe the possible outcomes of laboratory experiment.  As in Section \ref{bubbles} and \ref{ourbubble} it} includes a quantum picture of pocket universes at different levels of coarse graining.  As in Sections \ref{lambda}  and  \ref{st}  it provides a notion of a multiverse in which constants vary in a way that is different from pocket multiverses.

\subsection{How Specific Models Support  the General Conclusions}
\label{specmodels}
Sections V-VIII exhibited five quantum cosmological models that illustrate the different kinds of quantum multiverse that follow from an \HP's consisting  the no-boundary quantum state of  the universe  $\Psi$ and dynamical  theories $H$ based general relativity coupled to a single scalar field moving in different potentials.  These models are of course consistent with the general conclusions above as are the two-slit models in Section \ref{dh}.  But they also illustrate the following specific consequences:

$\bullet$ How pocket (bubble) multiverses  universes emerge from some dynamical theories \HP\ like the one in Section \ref{bubbles} but not from others like the ones in Sections \ref{homoiso} and \ref{ourbubble}.

$\bullet$. How the same theory \HP\ can predict multiverses at vastly different levels of coarse graining like the multiverse of true vacuum bubbles  in Section \ref{bubbles} and  the multiverse of possibilities for our bubble in Section \ref{ourbubble}.

$\bullet$. How to use appropriate coarse grainings implemented by summing amplitudes rather than probabilities to make manageable predictions for our observations even when the large scale structure of the universe is a  complex roiling sea of false vacuum bubbles in an ever expanding sea of false vacuum as in Section \ref{ourbubble}.

$\bullet$ How deterministic classical physics emerges from \HP\ as a quantum multiverse of histories with high probability for ones exhibiting correlations in time by deterministic laws like the Einstein equation as in  Section \ref{homoiso}.

$\bullet$ How a landscape potential with many minima leads to different possible values of the cosmological constant and how first person probabilities implement anthropic restrictions for values we will observe as in Section \ref{lambda}

$\bullet$ How a landscape of effective potentials leads to testable predictions for features of the CMB as in Section \ref{st}. 

 Quantum multiverses are what a fundamental theory \HP\ predicts. It is through them that we understand our universe.  As Weinberg wrote: ``Most advances in the history of science have been marked by discoveries about nature, but at certain turning points we have made discoveries about science itself''. The quantum multiverses of the universe are one of these turning points.

\acknowledgments

The author thanks Murray Gell-Mann, Stephen Hawking, Thomas Hertog, and Mark Srednicki  for discussions on the quantum mechanics of the universe over many years. He thanks Bernard Carr, George Ellis, and Thomas Hertog  for useful conversations, for critical readings of the manuscript, and for supplying relevant references. He thanks David Krakauer and Steven Benner for discussions on experimental evolution. He thanks Thomas Hertog for permission to reproduce Fig \ref{prior}. He thanks the Santa Fe Institute for supporting many productive visits there. This work was supported in part by the National Science Foundation under grant PHY15-04541.

 \appendix

\section{ A FAQ for Discussion}
\label{discussion}

There has been much discussion of cosmological multiverses by many authors. Some works known to the author (but not necessarily read carefully by the author) are \cite{CR1979,Wei05,Vil06,Car07,CE08,MH08,
Ste10, Ell11,VT11,Wil13,ES14,Ell14,Lin16}. This FAQ addresses  concerns that have been raised in some of these articles  about whether a quantum multiverse is falsifiable, testable, predictive, real, and  scientific. No claim is made to cover all of the issues that have been raised or to review all the discussion that has taken place.  This FAQ is separated from the main discussion because it reflects the opinions of the author more than the results of calculation. 

The quantum multiverse framework for prediction presented here is a synthesis of many elements. These include a quantum mechanics for the universe (DH), a model of observers, for example  the one in \ref{obsobs}, a theory  of the universe's quantum state  $(\Psi)$,  and a dynamical theory $(H)$  expressed in particular variables, perhaps defining a landscape of vacua, etc. 
\sk

{\it Is the notion of a quantum multiverse of the universe {\bf falsifiable?}} {\bf Yes.}  As in every other theory in physics this one can be falsified by falsifying any of the elements mentioned above that went into its construction. Take, for example DH. There is overwhelming evidence for quantum multiverses on laboratory scales. But there is little evidence that the same quantum mechanics (DH) can be extrapolated to the scales of cosmology as has been assumed in this paper.  Suppose quantum laboratory experiment  shows that DH is false on some intermediate scale. Then DH would be falsified. Similarly with the other elements of the predictive framework: a dynamical theory, a quantum state, a landscape for the variation of constants, etc. 
\sk

{\it Are quantum multiverses of the universe {\bf testable}? }{\bf Yes.} A theory like \HPs is testable by the (1st person)  probabilities it predicts  for the outcomes of observations. These are probabilities supplied by \HPs conditioned on a description of our observational situation\footnote{In various works we have called these {\it top-down} or {\it first person} probabilities.  For more on this see \cite{HH15b}.}. Ideally we would test the theory with situations where the predicted probabilities are near $1$. If a prediction with probability near $1$ does not occur the theory is falsified. But this luxury is seldom accessible in environmental sciences like geology, planet formation, biological evolution, human history, and cosmology. Rather we judge the success of a theory by its successful prediction of correlations among our present data as in the example in Section \ref{lambda} and Section \ref{st}.  (For more on criteria for success see, e.g. \cite{HH13}.) Hertog's prediction of the tensor/scalar ratio in the CMB described Section \ref{st} provides a clear example. If the observed ratio is much different from the predicted one then  either the landscape is wrong, the state is wrong, or the framework of quantum prediction is wrong. 
\sk

Something like a cosmological multiverses might be directly testable by experiment if a very large spacetime volume could be prepared with an initial quantum state from which galaxies, stars, life etc would emerge over billions of years.  This is quite beyond human powers at present. But the laws of the universe do not have to be such as to make it easy for some negligible bits of protoplasm  to test them directly on the scales that happen to be  accessible to them at the moment. 
\sk

{\it Are {\bf other histories in the multiverse of alternatives observable?}} {\bf No.} The different histories are exclusive alternatives. We are not outside the universe observing the whole ensemble, but rather inside  it  participating in the superposition.  A live Schr\"odinger cat does not observe a dead cat.  There are six alternative outcomes to the roll of a die each with a probability  of $1/6$. We observe the one face that comes up. We do not observe the alternative outcomes of that  one roll.
\sk

{\it Is the quantum multiverse of the universe {\bf real}?} {\bf Yes.} It is a tenet of many Everettian formulations of quantum theory for closed systems that all the histories in  a given multiverse are real  \cite{Wal12,mw10}.  Further, all the other multiverses  based on different variables with different levels of coarse graining are real. There are also formulations of DH in which only one history occurs in a multiverse of possible ones real \cite{GH11}.    For more of the author's  more nuanced views on what's real see \cite{Har06b,QU1}. 
\sk

\begin{figure}[t]
\includegraphics[height=6in]{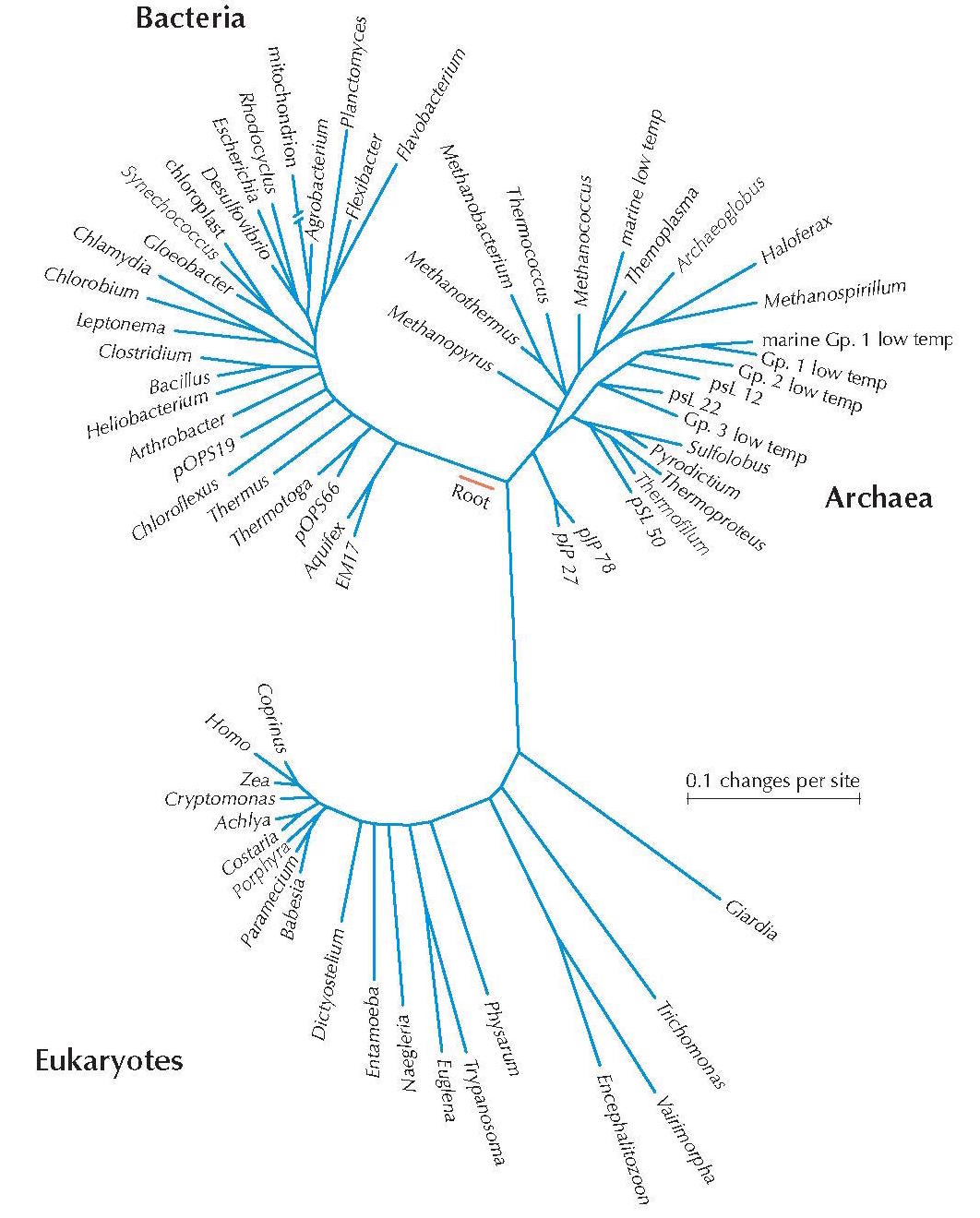}\hfill 
\caption{A single history of biological evolution in a coarse graining that only distinguishes certain genera \cite{Etree}. In this representation time roughly starts at `root' and proceeds outward. Different genera branch into others due to the accidents of biological evolution ---  mutation, recombination, genetic drift, extinction, etc. 
This is just {\it one} history of how the genera might have been formed. 
The evolutionary multiverse consists of all the different possible trees like this one representing different ways that the branching structure could have happened.  We do not directly observe any of this history. It's in the past.  Yet we believe in the multiverse of biological evolution because it explains regularities between species that we observe today. }
\label{evo}
\end{figure}
{\it Are there {\bf other areas of science} with similar issues?} {\bf Yes.}
 Biological evolution is concerned with the multiverse of different histories of the origin of  species. Individual histories can be illustrated by familiar tree diagrams like the example in Figure \ref{evo}.   Different evolutionary histories correspond to different branching trees. In principle the theory \HPs predicts probabilities for these various trees to occur but these are much beyond our ability to compute at present. We do not observe directly (`see')  any of the histories of  this multiverse. They are in our past. Yet we believe that biological evolution happened because it is explains regularities today.  For example plausible reconstructions of the evolutionary multiverse explain  the similarities between species today. Those similarities are consequences of frozen accidents of evolution --- chance events whose consequences proliferated. The theory of this multiverse is based on a number of elements:  Genetic variation by mutation, genetic drift, or recombination; a fitness landscape of ecological niches;  an initial condition of primordial DNA synthesis. This  is testable in a limited way by controlled laboratory experiment \cite{Lenski03,WK00}. Other historical sciences  such as geology, planet formation, human history  etc. provide similar examples of multiverses.  In cosmology we do not observe big bang  nucleosynthesis directly but we believe that it happened in the past   because of its successful prediction of  the correlations in the abundance of the elements  observed today.
 \sk
 
{\it Are quantum multiverses a {\bf departure from  laboratory physics?}} {\bf No}   Laboratory experiments are inside the universe not somehow outside it.  Laboratory experiments can therefore  be described by highly coarse grained alternative histories of the universe.  The highly successful Copenhagen formulation of quantum mechanics is not an alternative to DH   but an approximation to it that is appropriate for measurement situations \cite{Har91a}. Quantum cosmology is not in conflict with the successes in the laboratory.

  But  Copenhagen quantum mechanics must  be generalized beyond the 	the laboratory to apply the early universe where no measurements were being made and no observers were around to carry them out. DH, the generalization used here,  is not a departure from Copenhagen quantum mechanics where it applies but  rather a generalization of it to new domains of applicability. Some generalization is inevitable for cosmology and various ones been studied since the time of Everett. 
  
  Cosmology  is a historical science like geology, biological evolution, and human history.  Its aim is to use that data that exists now to reconstruct the quantum past in order to simplify the prediction of the future (e.g.\cite{Har98b}). We should not be surprised that the extension of observation to  new domains of phenomena require new extensions of existing theory.
  \sk

{\it Are quantum multiverses {\bf scientific.}} {\bf Yes, in the author's opinion:} As sketched above, in many areas of science one finds multiverses of the kind mentioned in the introduction --- an ensemble of possible  situations only one of which is observed by us. 

\section{A Little More DH}
\label{lm}

In this Appendix we repeat much of the qualitative discussion of the two slit models in Section \ref{dh} with more equations. That may helpful to some readers.

The inputs to the prediction of a quantum multiverse are the  Hamiltonian $H$ and quantum state  $|\Psi(t)\rangle$ of the particles in the boxes. This is a function of time in the Schr\"odinger picture in which we work. The state of the electron can also be described by a wave function in configuration space, viz.
\be
\Psi=\Psi(x,y,t)
\label{wvfn}
\ee
using a coordinate $x$ for the horizontal direction and $y$ for the vertical direction in the three boxes as shown in Figure \ref{2slit} and assuming symmetry in the perpendicular direction.
 We move back and forth between wave functions like \eqref{wvfn} and bras and kets like $|\Psi(t)\rangle$ as convenient. 

Histories can be represented by quantum branch state vectors constructed from \HP.   
For an example take the TSS model.   Denote the initial state at time $t_0$  by $\Psi(x,y,t_0)$. This is a product of wave packet in the $x$-direction $\phi(x,t_0)$  and a wave function $\psi_0(y,t_0)$ localized at the gun, viz.
\be
\label{fullstate}
\Psi(x,y,t_0) =\psi(y,t_0)\phi(x,t_0) 
\ee
This wave function evolves in time by the Schr\"odinger equation
\be
\label{schrod}
i \hbar \frac{\partial \Psi}{\partial t} = H \Psi.
\ee

We assume that   $\phi(x,t)$ is a narrow wave packet peaked to the left of the slits but moving to the right so as to reach the slits at time $t_s$ and the detecting screen at $t_d$. Thus, its progress in $x$ recapitulates evolution in time. 
After passing through the slits the wave function has the approximate form 
\begin{subequations}
\label{psiU}
\begin{align}
\Psi(x,y,t)&=\psi_U(y,t)\phi(x,t)  + \psi_L(y,t)\phi(x,t),  \quad t_s<t<t_d . \label{PSUa} \\
                 &\equiv  \label{PSUc}  \Psi_U(x,y,t) + \Psi_L(x,y,t).
\end{align}
\end{subequations}
Here, in the first term $\psi_U(y,t)$ is localized near the upper slit at time $t_s$ and spreads over a larger region of $y$ by the time $t_d$ that the electron hits the detecting screen. Similarly for the second term. The last line defines branch wave functions    $\Psi_U(x,y,t)$ and $\Psi_L(x,y,t)$ for the two histories in the set.

{\it Simple two-slit model (TSS):}  The electron starts localized at the gun. The alternative position intervals $Y$ at the further screen defines a coarse grained set  of alternative coarse-grained histories of the electron  between the times $t_0$ and $t_d$.  If $\{P_Y\}$ are a complete set of orthogonal projections on these intervals the branch state vectors for these  histories are  $|\Psi_Y(t_d)\rangle\equiv P_Y|\Psi(t_d)\rangle$.   The probability to arrive at $Y$  predicted from \HP\ is 
\be
\label{probY}
p(Y)=||P_Y|\Psi(t_d)\rangle||^2 = |||\Psi_Y(t_d)\rangle||^2 .
\ee

A finer grained set of histories for TSS would also specify whether the electron passed through the the upper slit $U$ or the lower slit $L$ on its way to a given interval $Y$. Naively one might expect  that the probabilities  $p_U(Y)$ and $p_L(Y)$ that the electron the upper or lower slit respectively and arrived  at $Y$ would be, from \eqref{PSUc},
\be
p_U(Y)=||P_Y|\Psi_U(t_d)\rangle||^2, \quad p_L(Y)=||P_Y|\Psi_L(t_d)\rangle||^2 
\label{branch}
\ee  
Then we would have from \eqref{PSUc}  and \eqref{probY}
\be
\label{inconsist} 
p(Y)=||P_Y|\Psi_U(t_d)\rangle + P_Y|\Psi_L(t_d\rangle||^2  =||P_Y\ |\Psi_U(t_d)\rangle||^2 + ||P_Y|\Psi_U(t_d)\rangle|||^2
\ee
This is false unless  the two branches $P_Y|\Psi_U(t_d)\rangle$ and  $P_Y|\Psi_L(t_d\rangle$ are orthogonal so that the quantum interference between them vanishes. That is,  it is inconsistent  unless the set of histories decoheres.
\be
\label{decohcond}
\langle \Psi_U(t_d)|P_Y |\Psi_L (t_d)\rangle \approx 0
\ee
 That won't be the case in TSS. The two histories interfere as shown by  interference pattern that is a characteristic feature of the two slit experiment. TSS does not have a multiverse of this kind.

{\it Two-slit model with gas (TSG):}
The TSG model contains a gas of particles that interact weakly with the electron ---  an example of an {\it environment}. The initial wave function is
\be
\label{fullstatewg}
\Psi(x,y,t_0) =\psi(y,t_0)\phi(x,t_0)\chi(t_0) , \quad t_0<t<t_s.
\ee
where $\chi(t_0)$ is the initial state of all the particles in the gas. After passing through the slits this becomes
\begin{subequations}
\label{psiUwgas}
\begin{align}
\Psi(x,y,t)&=\psi_U(y,t)\phi(x,t)\chi_U(t)  + \psi_L(y,t)\phi(x,t)\chi_L(t),  \quad t_s<t<t_d . \label{PSUga} \\
                 &\equiv  \label{PSUgc}  \Psi_U(x,y,t) + \Psi_L(x,y,t). 
\end{align}
\end{subequations}
where $\chi_U(t)$ and $\chi_L(t)$ denote the state of the gas particles that have scattered from the region of the upper and lower slit respectively. Then  $\Psi_U(x,y,t_d)$ and $\Psi_L(x, y,t_d)$ are the branch wave functions for the two histories that the electron arrived in $Y$ at time  $t_d$ after passing through either the upper or lower slit. 

In the simple two-slit model (TSS) the set of alternative histories describing which slit ($U$ or $L$) the electron went through on its way to arrive at $Y$ on the further screen did not decohere as required for a quantum multiverse. That was because the states representing the two histories were not orthogonal.  However, because of the gas in (TSG) the two histories do decohere. The decoherence condition \eqref{decohcond} is satisfied because the state of the gas scattered from the upper slit $|\chi_U\rangle$ is orthogonal to the state of the gas scattered from the lower slit $|\chi_L\rangle$ if enough particles scatter. 

To see that gas ensures this orthogonality consider a single  gas particle $a$ whose initial state is $|\chi_a\rangle$ that scatters near the upper slit will be in a final state $S_U |\chi_a\rangle$ where $S_U$ is the S-matrix for scattering near the  upper slit. Similarly for the lower slit with $S_L|\chi_a\rangle$  If $N$ gas particles scatter the overlap between $|\chi_U\rangle $ and $|\chi_L\rangle$ will be proportional to  $N$  inner products of the form
\be
\label{overlap} 
\prod_{a=1}^N |\langle\chi_a| S_U^\dagger S_L|\chi_a\rangle|. 
\ee
Since the two final state vectors $S_U |\chi_a\rangle$ and $S_L|\chi_a\rangle$  are different each individual product has a magnitude less than $1$. The product of a very large number of such products will be near zero implying decoherence.

Thus for TSG we have exhibited two quantum multiverses  at different levels of coarse graining. The first one (just $Y$) is a coarse graining of the second ($Y$ and $S$). The first ignores the gas, the second follows it. 

\vskip .3in

\end{document}